\documentclass[preprint,preprintnumbers,amsmath,amssymb,superscriptaddress,nofootinbib]{revtex4}

\usepackage{graphicx}
\usepackage{psfrag}
\usepackage{axodraw}
\usepackage{subfigure}
\usepackage{color}
 \newcommand{\be}{\begin{equation}}
 \newcommand{\ee}{\end{equation}}
 \def\bea{\begin{eqnarray}}
 \def\eea{\end{eqnarray}}
 \newcommand{\f}[2]{\frac{#1}{#2}}

\begin{document}

\author{Eduard Mass{\'o}}
\author{Javier Redondo}
\affiliation{Grup de F{\'\i}sica Te{\`o}rica and Institut de F{\'\i}sica d'Altes Energies
Universitat Aut{\`o}noma de Barcelona 08193 Bellaterra, Barcelona, Spain}

\title{Axion  results: what is new?\footnote{Based on two talks given at the International Workshop
"The dark side of the Universe", Madrid, June 2006: "Evading astrophysical bounds on axion-like
particles in paraphoton models" by J. Redondo and "Axion  results: what is new?" by E. Masso. To be
published in the Proceedings. }}


\begin{abstract}
The PVLAS collaboration has obtained results that may be interpreted in terms of a light axion-like
particle, while the CAST collaboration has not found any signal of such particles. Moreover, the
PVLAS results are in gross contradiction with astrophysical bounds. We develop a particle physics
model with two paraphotons and with a low energy scale in which these apparent inconsistencies are
circumvented.
\end{abstract}

\maketitle


\section{Introduction and motivation}

Particle physics theories that go beyond the standard $SU(3)\times SU(2)\times U(1)$ model  have
usually new global symmetries. When one of these global symmetries is spontaneously broken we get a
Goldstone  or a pseudo Goldstone boson. In general the new particle  $\phi$ will be light (or
massless) and will couple to two photons. Examples are extensions of the standard model involving
breaking of a family symmetry, or lepton number symmetry, or string theories \cite{Svrcek:2006yi}.

Depending on the parity of the particle the two photon coupling is described by the lagrangian
\begin{equation}
{\cal L}_{\phi\gamma\gamma}=\frac{1}{8M} \epsilon_{\mu\nu\rho\sigma} F^{\mu\nu} F^{\rho\sigma}\,
\phi
 \label{masso_L}
\end{equation}
for a pseudoscalar, while for a scalar we would have
\begin{equation}
{\cal L}'_{\phi\gamma\gamma}=\frac{1}{4M}
 F_{\mu\nu}
F^{\mu\nu}\, \phi \label{masso_LS}
\end{equation}
We will denote these new hypothetical light particles by $\phi$, and refer to them as axion-like
particles (ALPs) \cite{Jaeckel}, both for the scalar and pseudoscalar case.

ALP physics has received a lot of attention lately
\cite{Jaeckel,nature,Masso:2005ym,recentwork,abel,Masso:2006gc} because ALPs are a plausible
explanation of the recent results of the Polarization of Vacuum with Laser (PVLAS) collaboration
\cite{ZavattiniEM}. They observe a rotation of the plane of the polarization of a laser when
propagating in a magnetic field. The result can be interpreted as production of an ALP with a mass
 \begin{equation}
m \sim 10^{-3}\ {\rm eV}  \label{masso_m_pvlas}
\end{equation}
and a scale interaction
\begin{equation}
M \sim 4 \times 10^{5}\ {\rm GeV}  \label{masso_pvlas}
\end{equation}

The story is not finished since the coupling (\ref{masso_L}) or (\ref{masso_LS}) let $\phi$
particles to be produced copiously in the center of our Sun or other stars like red giants. The
production mechanism is $\gamma \rightarrow \phi$ in the electromagnetic field of a nucleus or an
electron of the star core (this is  analogous to the observed Primakoff effect $\gamma \rightarrow
\pi^0$ in the electromagnetic field of a nucleus). The value (\ref{masso_pvlas}) is low enough so
that the produced ALP escapes the star with no interactions. This would mean a quite large
luminosity ${\cal L}_\phi$ in this exotic channel. The value (\ref{masso_pvlas}) implies for the
Sun
 \begin{equation}
{\cal L}_\phi \sim 10^6 \,  {\cal L}_\odot
\end{equation}
Of course, this would be a disaster for the solar evolution.

In these Proceedings we would like to summarize some models that we have developed and where the
puzzle is solved. In our models, the coupling  (\ref{masso_L}) or (\ref{masso_LS}) is valid at the
very low energies of the PVLAS experiment. However, it gets modified when going to the conditions
of the stellar interior. Obviously we look for models where the effective coupling is strongly
diminished in the stellar environment. The lagrangians (\ref{masso_L}) and (\ref{masso_LS}) are
five-dimensional operators so that if we wish to get a modification of the   coupling we assume a
new energy scale of energy much less than the   typical stellar interiors, ${\cal O}$(1 keV). Here
we will only look at modifications to the $\phi\gamma\gamma$ coupling due to the relatively high
temperatures of the Sun. There are other parameters, like the environment density, that also could
affect the coupling. This is studied in   \cite{Jaeckel}.

Solving in our way the apparent problem of the PVLAS results when examining its astrophysical
consequences, has an additional bonus. There is another puzzle concerning ALPs, that originates in
the result of experiment run by the CAST collaboration \cite{ZioutasEM}.  The CAST experiment is a
helioscope  \cite{SikivieEM}, which expects to detect the solar flux of $\phi$ particles by means
of their conversion in X-rays in a cavity with a strong magnetic field. The null CAST result
implies a limit on $M$,
\begin{equation}
M > 0.9 \times 10^{10}  \ {\rm GeV}   \label{masso_cast}
\end{equation}
The puzzle is clear, (\ref{masso_cast}) and (\ref{masso_pvlas}) are in gross contradiction. Our
models solve also this contradiction, since if we are able to diminish the ALP production in the
Sun, then (\ref{masso_cast})  is no longer valid since the CAST bound assumes a standard emission
of ALPs.

\newpage
\section{Paraphoton Models}

Our strategy to lower the novel particle emission from stellar environments is to provide a
particular structure to the interactions (\ref{masso_L}) or (\ref{masso_LS}): a triangle diagram
with a new fermion running in the loop (shown in Fig.\ref{triangle}). The matching between this
diagram and eq.(\ref{masso_L}) gives
\begin{equation}
\frac{1}{M} = \frac{\alpha}{\pi} \frac{Q_f^2}{v} \label{triangle}
\end{equation}

where $Q_f$ is the electric charge of $f$ and $v$ is a function of $m_f$ and $m_\phi$ if $\phi$ has
a scalar or pseudoscalar coupling but can be an completely independent energy scale if $\phi$ is a
Goldstone boson.
\begin{figure}[h]\label{triangle}
  \includegraphics[width=5cm]{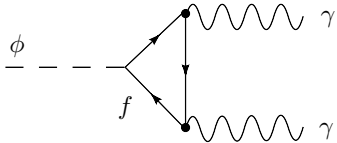}\vspace{-.4cm}
  \caption{Triangle diagram responsible for the PVLAS signal in our model.}
\end{figure}

As we want $v$ to be a low energy scale we shall consider models in which $Q_f$ is very small. This
can be naturally achieved in paraphoton models \cite{Holdom:1985ag} which can be considered as
extensions of QED. The simplest model assumes the addition of new gauge $U(1)_1$ symmetry to the
usual electromagnetic one which we will call $U(1)_0$. As a consequence, the theory has a new gauge
boson, $A_1$, called paraphoton. Two further ingredients of a typical paraphoton model are fermions
that can have electric and/or para-charge and mixing terms $\epsilon F_0^{\mu\nu} F_{1\mu\nu}$
between the field strengths $F_i^{\mu\nu}$ of the gauge bosons. The later are allowed by the
combined symmetry $U(1)_0\times U(1)_1 $ and thus must be present in a renormalizable lagrangian.
Furthermore, even if they are not included at the beginning in the theory they are radiatively
generated by massive fermions charged under both $U(1)$'s. In what follows we are going to consider
that this is the case, so small values of $\epsilon$ are natural on the basis of their radiative
origin.

The important point is that these mixing terms act as if there where new contributions to the
fermion charges $Q_i$. This can be explained in several ways. In \cite{Masso:2005ym} we use one of
them first presented in \cite{Holdom:1985ag} that consists in diagonalizing together the kinetic
and mass parts of the lagrangian for the gauge bosons.  Let us develop here a simpler (though less
formal) method of calculation.

In the LHS of Fig.\ref{efef-mass} we see the interaction of a para-fermion $f$ belonging to the
representation $(0,Q_1^f)$ with an electron $\in (Q_0^e,0)$. We consider the general case in which
both gauge bosons are massive. The value of this amplitude is
\be {\cal A}=(e_0Q^e_0 j^e_\mu) \f{i}{q^2-m_0^2} (\epsilon i q^2) \f{i}{q^2-m_1^2
}(e_1Q^f_1j^{f\mu}) \label{efef-mass-amplitude} \ee

($q$ is the momentum carried by the bosons, $e_0,e_1$ are the coupling constants and
$j_\mu^e,j_\mu^f$ the currents of electrons and $f$ particles) which we can decompose using
\be
\f{-iq^2}{(q^2-m_0^2)(q^2-m_1^2)}=\f{m_0^2}{m_1^2-m_0^2}\f{i}{q^2-m_0^2}+\f{m_1^2}{m_0^2-m_1^2}\f{i}{q^2-m_1^2}
\ee

to realize that this amplitude is completely equivalent to the sum of two single boson exchange
diagrams as shown in Fig.\ref{efef-mass} which require that we assign an $\epsilon$-charge to $f$
and a $\epsilon$-para-charge to the electron $e$:
\begin{figure}[t] \label{efef-mass}
  \includegraphics[width=16cm]{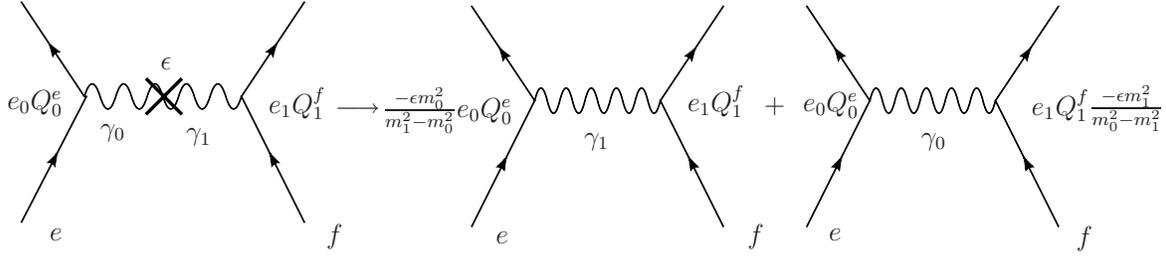}
  \caption{Mixing between massive photons is equivalent to new $\epsilon$-charges for the fermions.}
\end{figure}
\be Q_0^f=\epsilon Q_1^f \f{m_0^2}{m_1^2-m_0^2}\f{e_1}{e_0} \hspace{1cm} Q_1^e=\epsilon Q_0^e
\f{m_1^2}{m_0^2-m_1^2}\f{e_0}{e_1}  \label{efef-charges}\ee
These charges are in agreement with eq.(15) of \cite{Masso:2005ym} which was derived by a different
method. Note in particular that if a boson is massless and the other massive, only particles that
coupled to the massless boson acquire new charges. A final remark is worth, namely that we are
adjusting the charges by comparing diagrams in perturbation theory and clearly when the photons are
degenerate in mass the charges diverge so our formulae do not hold for the case $m_0=m_1\neq 0$.
Interestingly enough they can be useful for the case in which the two bosons are massless because
we can fulfill the perturbative condition by performing first the limit $m_0\rightarrow 0$ and
after $m_1\rightarrow 0$ or viceversa. The result is shocking at first sight: the two orderings
give different assignments of charge! To understand that this is not an inconsistency let us show
the value of the $ef\rightarrow ef$ amplitude in this case:
\be {\cal A}=(e_0 Q^e_0 j^e_\mu) \f{i}{q^2} (\epsilon i q^2) \f{i}{q^2 }(e_1Q^f_1j^{f\mu})=
(e_0Q^e_0 j^e_\mu)\f{-i\epsilon}{q^2}(e_1Q^f_1j^{f\mu}) \ee
We see that this amplitude could be attributed to diagram with exchange of a $\gamma_0$ an a new
$\epsilon$-sized $Q_0^f$
\be Q_0^f= -\epsilon Q_1^f e_1/e_0 \label{Q-vacuum}\ee
or to a diagram with exchange of $\gamma_1$ and a $\epsilon$-sized $Q_1^e$:
\be Q_1^e=-\epsilon Q_0^e e_0/e_1 \ee
Indeed, generally we can put a combination of both such that the sum remains the same. This would
correspond to take the limit $m_0^2=\alpha m_1^2\rightarrow 0$ in eq.(\ref{efef-charges}) (with
$|\alpha -1| \gg \epsilon$) which gives:
\be Q_0^f=\epsilon Q_1^f \f{1}{\alpha-1}\f{e_1}{e_0} \hspace{1cm} Q_1^e=\epsilon Q_0^e
\f{\alpha}{1-\alpha}\f{e_0}{e_1} \ee

This freedom in the assignment of charges can be traced back to the freedom we have to rotate the
basis $\{A_0, A_1\}$ because the photons are degenerated in mass as we comment in
\cite{Masso:2005ym}. We find then that in this case the $\epsilon$-charge assignments are
convention dependent.

Now that we can calculate easily charges arising from mixing terms in paraphoton models we are
going to apply our knowledge to explain our solution to the PVLAS-CAST apparent inconsistency.

\subsection{A model with two paraphotons solving the PVLAS-CAST inconsistency}

The motivation of our model comes from the high $q^2$ behavior of the LHS diagram in
Fig.\ref{efef-mass} shown in (\ref{efef-mass-amplitude}). As we go higher in $q^2$ the dependence
on the paraphoton mass $m_1$ becomes smaller. Then we can consider that in adition to the amplitude
of Fig.\ref{efef-mass} we have other diagram with a different paraphoton $\gamma_2$ with
\textit{different mass} $m_2$ that, having opposite sign, cancels the first one at high $q^2$
(corresponding to the typical stellar environment) but leaves a finite contribution at low $q^2$
where the PVLAS experiment takes place.
\begin{figure}[h] \label{efef-para}
  \includegraphics[width=15cm]{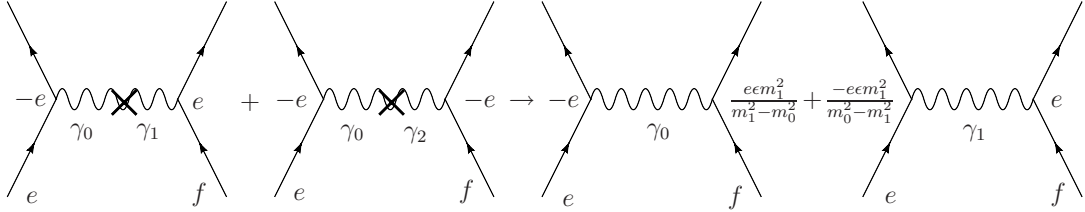}
  \caption{$e f \rightarrow ef$ from which we can infer the charges in our paraphoton model.
  We have set $e_0=e_1=e_2=e$, $Q_0^e=-1$, $Q_1^f=-Q_2^f=1$. The two LHS diagrams are equivalent
  to the two RHS. The diagram of the $\gamma_0$ exchange recieves contributions from $\gamma_0-\gamma_1$
  and $\gamma_0-\gamma_2$ mixing diagrams that cancel at high $m_0$, while the later is due only
  to $\gamma_0-\gamma_1$.}
\end{figure}

We consider then a model with two paraphotons in which the local symmetry is $U(1)_0\otimes U(1)_1
\otimes U(1)_2$. The condition that we have to impose for the two diagrams to cancel at high
momentum transfer is $\epsilon_{01}e_1Q_1^f-\epsilon_{02}e_2Q_2^f=0$ as can be deduced from
Fig.\ref{efef-para}. We can set $e_0=e_1=e_2$ for simplicity. Very massive fermions $F$ in the
representations $(Q_0^F,Q^F,Q^F)$ will produce naturally $\epsilon_{01}=\epsilon_{02}$ and we can
choose $Q_1^f=-Q_2^f=1$ for the light para-fermion $f$. The second condition is that at the low
momentum transfer of the PVLAS experiment ($q^2 \sim 10^{-6}$ eV$^2$) $Q_0^f$ should be finite and
preferably maximum. Note from \eqref{efef-charges} that massive paraphotons imply $Q_0^f=0$ when
photons are in vacuum ($m_0=0$), so we need one paraphoton to be massless\footnote{Strictly
speaking, we must require that the paraphoton has a mass smaller than the uncertainty of the
momentum $p$ of the photons in the PVLAS experiment coming from the Heisenberg uncertainty
principle $\Delta x\Delta p>1$.}. Accordingly we choose $m_1=\mu\neq 0$ and $m_2=0$. We must,
however, note that in the sun the dispersion relation of photons resembles that of a massive
particle with mass equal to the plasma frequency so $m_0=\omega_P \sim 4\pi \alpha n_e/m_e$
($\alpha\sim1/137$, while $n_e$ and $m_e$ are the electron density and mass, respectively). Then we
find that the electric charge of $f$ depends on the environment where it is probed, and from
\eqref{Q-vacuum} and \eqref{efef-charges} we get:
\be Q_0^f(PVLAS) = \epsilon\hspace{1cm};\hspace{1cm} Q_0^f(Sun) \sim -\epsilon\f{\mu^2}{\omega_P^2}
\ee
We reach our goal of having a decrease of the electric charge of $f$ and consequently of the novel
particle emission from the sun by requiring $\mu/\omega_P$ to be small enough.

\subsection{Astrophysical bounds evaded}

We now discuss the consequences of our model. The PVLAS experiment is in vacuum, so $f$ has an
effective electric charge $Q^f_0=\epsilon$, which from (\ref{triangle}) has to be
\begin{equation}
\epsilon^2 \simeq   10^{-12}\, \frac{v}{\rm eV}
 \label{N1}
\end{equation}

Concerning the astrophysical constraints \cite{Raffelt:1996wa} we should first look for the
relevant production processes of the exotic particles in our model. We notice that the amplitude
for the Primakoff effect $\gamma Z \rightarrow \phi Z$ is of order
$\epsilon^2(\mu^2/\omega_P^2)^2$. But there are production processes with amplitudes of order
$\epsilon(\mu^2/\omega_P^2)$ which will be more effective. The most efficient is plasmon decay
$\gamma^* \rightarrow \bar f f$. Energy loss arguments in horizontal-branch (HB) stars
\cite{Davidson:1991si} limits $Q_0^f < 2\times 10^{-14}$, which translates in our model into the
bound
\begin{equation}
\epsilon\  \frac{\mu^2}{ \rm eV^2} <   4\times 10^{-8}
 \label{N2}
\end{equation}
(we have introduced $\omega_P \simeq 2$ keV in a typical HB core). But equations (\ref{N1}) and
(\ref{N2}) do not fully determine the parameters of our model. Together they imply the constraint
\begin{equation}
v\, \mu^4 <  (\, 0.4 \ {\rm eV})^5
\end{equation}

We can now make explicit one of our main results. In the reasonable case that $v$ and $\mu$ are not
too different, we wee that \textit{the new physics scale is in the sub eV range}.

On the other hand, the CAST telescope is able to detect $\phi$'s with energies within $1$ and $14$
keV. In our model, $f$'s and paraphotons are emitted from the Sun, but we should care about $\phi$
production. We consider three possibilities.

A) $\phi$ is a fundamental particle. Production takes place mainly through plasmon decay
$\gamma^*\rightarrow \bar{f}f \phi$. The $\phi$-flux is suppressed, but, most importantly, the
average $\phi$ energy is much less than $\omega_P \simeq.3$ keV, the solar plasmon mass. The
spectrum then will be below the present CAST energy window.

B) $\phi$ is a composite $\bar{f} f$ particle confined by new strong confining forces. The final
products of plasmon decay would be a cascade of $\phi$'s and other resonances which again would not
have enough energy to be detected by CAST \cite{citaK}.

C) $\phi$ is a positronium-like bound state of $\bar{f} f$, with paraphotons providing the binding
force. As the binding energy should be small, ALPs are not produced in the solar plasma.

A final constraint should bother us. In vacuum, $\gamma_1$ couples to electrons with a strength
$\epsilon$ and a range $\mu^{-1}$ and this interaction is limited by Cavendish-type experiments
\cite{Bartlett:1988yy}.

\begin{figure}[h]
 \includegraphics[width=9cm]{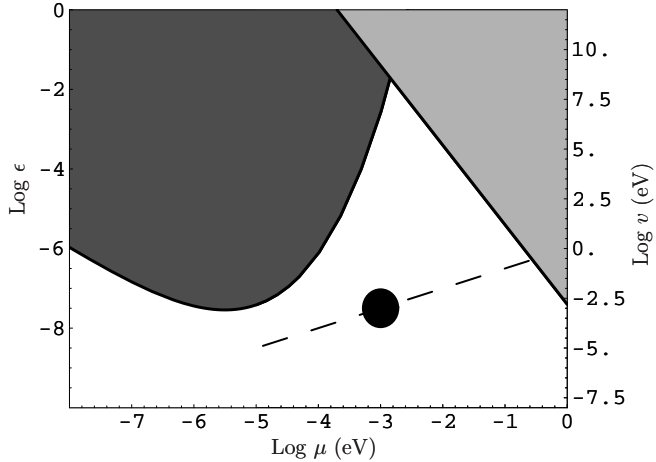}
 \caption{Constraints on the parameters of our model. The black
 area is excluded  by Cavendish-type experiments,
 and the grey area by the astrophysical constraint
 (\ref{N2}). The dashed line corresponds to $v=\mu$, and the dot to
 $v=\mu\simeq 1$ meV.
   \label{fig3} }
\end{figure}

Finally, in Fig.(\ref{fig3}) we show all these limits. In the ordinates we can see both $\epsilon$
and $v$, since we assume they are related by (\ref{N1}). We find out that there is wide room for
the parameters of our model, even in the natural line $v=\mu$ or further, in the preferred point
$v=m_f=m_\phi\sim$ meV.

\newpage

To conclude let us say that the PVLAS-CAST puzzle has recently received very much attention. We
have presented a model with two paraphotons and a new fermion $f$ living all at a low energy scale,
below eV, in which the apparent discrepancy is safely circumvented. Let us also mention that
recently our model has been justified in the context of string theory \cite{abel}.


\section{Acknowledgments}
We acknowledge support by the projects FPA2005-05904 (CICYT) and 2005SGR00916 (DURSI).



\bibliographystyle{aipproc}   


\end{document}